\documentclass{article}
\usepackage{arxiv}
\usepackage{graphicx}
\usepackage[nolist,printonlyused]{acronym}
\usepackage{amsmath,empheq,amsthm,bm,tabularx}
\usepackage[hidelinks]{hyperref}
\usepackage{booktabs}

\graphicspath{{./graphics/}}
\usepackage{cite}
\usepackage{tikz}

\title{Circular economy meets building automation}

\author{
  Hanmin Cai \\
  Urban Energy Systems Laboratory \\
  Empa, D\"{u}bendorf, Switzerland \\
  \texttt{hanmin.cai@empa.ch}
}

\begin{document}
\maketitle
\footnote{
This work has been accepted for publication in a CISBAT 2023 special issue of IOP's JOURNAL OF PHYSICS Conference Series. 
}
    \begin{abstract}
This paper demonstrates the concept of reusing discarded smartphones to connect the end-of-life of e-wastes with the start-of-life of smart buildings. Two control-related and one communication-related case studies have been conducted experimentally to evaluate the applicability. Diverse controlled systems, control tasks, and algorithms have been considered. In addition, the sufficiency of communication with external agents has been quantified. The proof-of-concept experiments indicate the technical feasibility and applicability to typical tasks with satisfactory performance. As the capabilities of smartphones improve over time, higher computing performance and lower communication latency can be expected, which enhances the prospect of the proposed reuse concept. 

    \end{abstract}

	\section{Introduction}\label{Intro}
	The circularity within the building sector can be enhanced by examining the e-wastes that are largely neglected so far.
On the one hand, ongoing digitization in the energy sector intensifies the demand for computing power, i.e., programmable logical controllers (PLCs) \cite{beckhoff} are commonly used for automation in buildings. On the other hand, many smartphones are mainly disposed of to extract valuable metals \cite{rizos2019identifying}. This paper contributes to bridging this gap by investigating the reuse of discarded smartphones to connect the end-of-use of smartphones with the start-of-life of smart buildings. 

Currently, many smartphones turn into e-wastes when they are outdated or have battery/screen malfunction \cite{rizos2019identifying}. However, their central processing units (CPU) and random-access memory (RAM) may remain intact. Therefore, they are potential neglected resources to perform building energy management tasks. Once connected to the internet and power sources, they can help to avoid manufacturing new micro-controllers and to reduce the overall carbon footprint. Although the reuse of outdated desktops has been investigated in lab facilities \cite{syslab}, the reuse of smartphones for building energy management systems has not been systematically investigated in experiments. Key concerns include the timely execution of control algorithms and effective communication with external devices. While the former impacts occupants' comfort at the building level, the latter impacts agent-based coordination among buildings \cite{cai2020agent} and attainable ancillary service options \cite{swissgrid}.

To systematically assess the technical feasibility of such a concept, several proof-of-concept experiments are provided in this paper, considering diverse controlled systems, control tasks, and control algorithms with different levels of computation and communication loads.

	\section{Methodology}\label{sec:Method}
	Two control algorithms are considered to reflect the complexity of control-related case studies, namely model-based predictive control and data-driven predictive control. They differ in terms of communication and computational loads. Both control strategies are key to enable optimal building-level operation and district-level coordinated building energy management. The rest of the section provides a brief summary.

The main principle of \ac{MPC} is recursively solving an \ac{OCP} over a horizon considering the future response of the system. While the context-specific constraints and cost functions are elaborated in section \ref{sec:CaseStudy}, a generic formulation is given as follows:
\begin{align}\label{eq:MPC}
\underset{\mathbf{u},\mathbf{x},\mathbf{y}}{\textup{minimize}}\quad
& \sum_{k=0}^{N-1}\left(\left\|\mathbf{y}_k-\mathbf{r}_{t+k}\right\|_\mathbf{Q}^2 +\left\|\mathbf{u}_{k}\right\|_\mathbf{Q'}^2\right) \nonumber \\
\text{subject to\quad}
& \mathbf{x}_{k+1}=\mathbf{A}\mathbf{x}_k+\mathbf{B}\mathbf{u}_k, \;\forall k \in \{0,  \ldots, N-1 \},\nonumber \\
& \mathbf{y}_k=\mathbf{C}\mathbf{x}_k+\mathbf{D}\mathbf{u}_k, \; \forall k \in \{0,  \ldots, N-1 \}, \nonumber \\
& \mathbf{x}_0=\hat{\mathbf{x}}_t, \\
& \mathbf{u}_k\in \mathcal{U},\; \forall k \in \{0,  \ldots, N-1 \}, \nonumber \\
& \mathbf{y}_k\in\mathcal{Y}, \; \forall k \in \{0,  \ldots, N-1 \},\nonumber
\end{align}
where $N$ is the time horizon, $\mathbf{u}$, $\mathbf{x}$, $\mathbf{y}$ are the vectors of decision variables, $\mathbf{r}$ is the reference for tracking, $t$ is the time stamp, $\mathbf{Q}$ and $\mathbf{Q'}$ denote the weighting matrices for input and output costs, $\mathcal{U}$ and $\mathcal{Y}$ are the input and output constraint sets, and $\hat{\mathbf{x}}_t$ is the estimated state at time $t$. 
By varying $\mathbf{Q}$ and $\mathbf{Q'}$, \autoref{eq:MPC} can be customized to represent constrained energy planning and reference tracking tasks.
In addition, to mitigate potential numerical instability when estimating $\hat{\mathbf{x}}_t$, a Kalman filter with Joseph formulation \cite{bucy2005filtering} is used, which is given as follows:
\begin{equation}
    \mathbf{P}^+ = (\mathbf{I}-\mathbf{K}\mathbf{H})\mathbf{P}^-(\mathbf{I}-\mathbf{K}\mathbf{H})^\text{T}+\mathbf{K}\mathbf{R}\mathbf{K}^\text{T},
\end{equation}
where $\mathbf{I}$ is the identity matrix, $\mathbf{K}$ is the gain, $\mathbf{H}$ is the
measurement mapping matrix, $\mathbf{R}$ is the measurement noise covariance matrix, and $\mathbf{P}^-$, $\mathbf{P}^+$ are the prior and
post measurement update estimation error covariance matrices, respectively.

Regarding data-driven control, \ac{SMM-PC} \cite{yin2021maximum} is considered. In brief, it maximizes the conditional probability of observing the predicted output trajectory and the measured past outputs to improve the combination of offline trajectories. Therefore, it differs from the previously mentioned \ac{MPC} in terms of real-time data extraction, the need for state estimation and the \ac{OCP}.

	\section{Case studies}\label{sec:CaseStudy}
	To systematically evaluate the applicability of the reuse concept in practice, diverse controlled systems, control tasks, and algorithms were considered in the experiments. This section includes a description of the general experimental setup that supports the case studies, followed by a detailed explanation of each case.

\subsection{General experimental setup}
\begin{figure}[htbp]
    \centering
    \includegraphics[width =.5\textwidth]{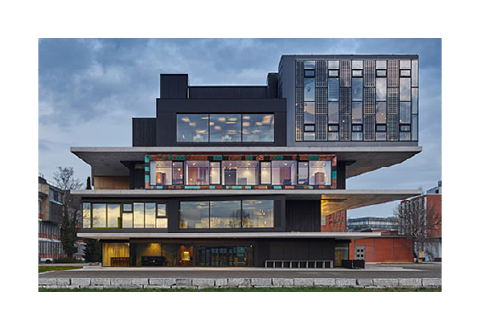}
    \caption{The NEST building in D\"ubendorf, Switzerland. Copyright @ Empa.}\label{fig:nest}
\end{figure}

\begin{figure}[htbp]
    \centering
    \includegraphics[width =\textwidth]{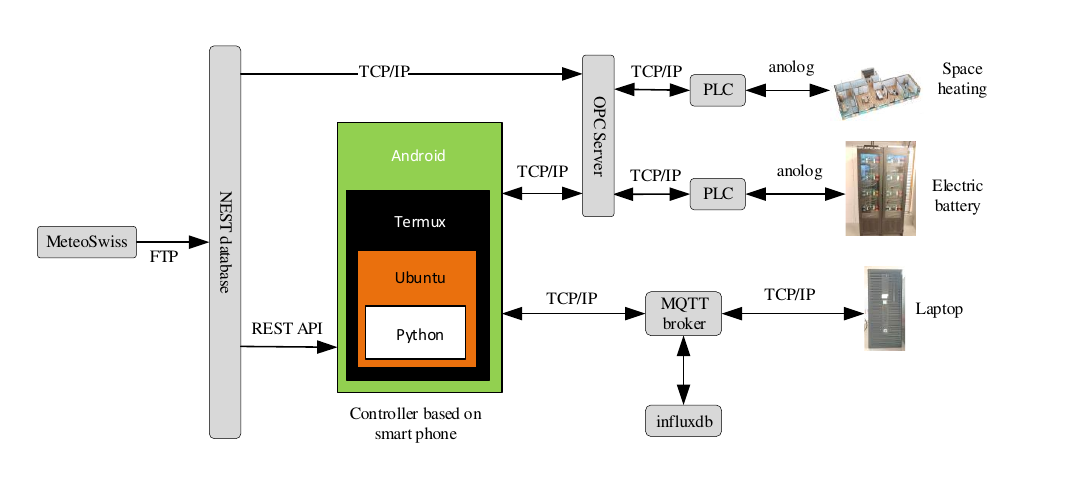}
    \caption{Software and information exchange setup for the experiments. The images of experimental facilities are sourced from \cite{nest}.}\label{fig:toolchain}
\end{figure}

The overall setup is illustrated in \autoref{fig:nest} and \autoref{fig:toolchain}. The open-source package \textit{Termux} \cite{termux}\footnote{The mention of this software does not constitute an endorsement of the product, and the authors are not affiliated with the software developers.} was used to enable smartphones to function as a Linux machine and host Python installation. \Ac{SSH} was used for remote/headless access to facilitate reusing discarded smartphones with cracked screens. The communication among devices was enabled by \ac{MQTT} \cite{Light2017} in the communication latency assessment case study. Real-time information exchange in the control-related case studies was facilitated with the REST API and an OPC server. 
\subsection{Applicability to control tasks}
\begin{figure}[htbp]
    \centering
    \includegraphics[width =0.7\textwidth]{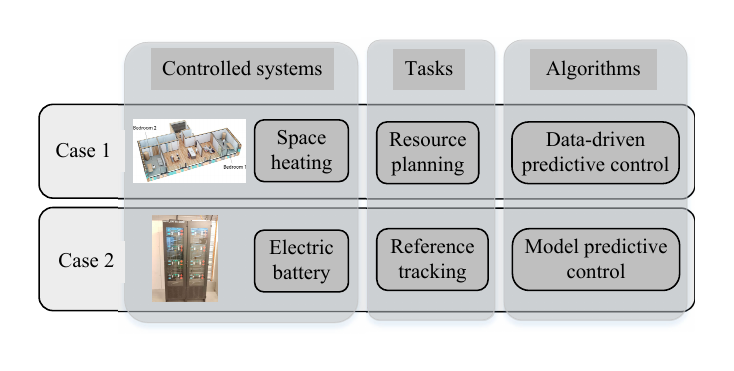}
    \caption{Control-related case studies composed of diverse controlled systems, control tasks and algorithms.}\label{fig:case_study_setup}
\end{figure}

To reflect the heterogeneous controlled systems in the built environment, both space heating and stationary electric battery were included. Specifically, optimal space heating power scheduling was considered and the reference tracking task was examined for the stationary electric battery, e.g., necessary concerning ancillary service provision \cite{swissgrid}. 
Both the classical \ac{MPC} and data-driven control \ac{SMM-PC} presented in the previous section were assessed. 
This diverse setup is further illustrated in \autoref{fig:case_study_setup}.

More concretely, \textit{Case 1} concerns minimizing the heating energy consumption of a bedroom of a three-room apartment while respecting user-defined temperature limits. The input and output constraint sets of the \ac{OCP} include heating power capacity and room temperature limits. The control decisions concerned heating power, which was dissipated through a ceiling heating system. The thermostat in the room was remotely controlled by manipulating the setpoints, which influenced the valve opening/closing to regulate the heating flow indirectly. Continuous control decisions were translated into sequences of binary valve positions using a pulse-width modulation (PWM) strategy. In \textit{Case 2}, a Lithium-ion electric battery was controlled to track an artificial sinusoidal reference signal with its \ac{SOC}. The input was constrained by charging and discharging power limits.
In both cases, the sampling time was $15$ minutes.

\subsection{Applicability to communication tasks}
\begin{figure}[htbp]
    \centering
    \includegraphics[width =\textwidth]{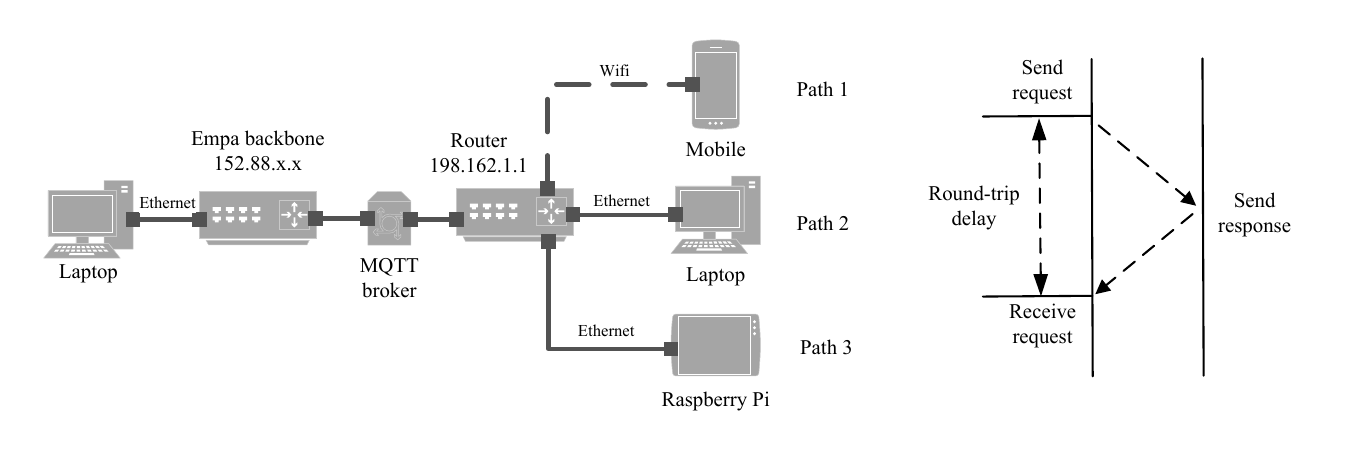}
    \caption{Setup of case studies for communication delay identification. The figure on the left shows the network configuration of three information exchange paths. The figure on the right illustrates the calculation of round-trip delays.}\label{fig:com_delay_test}
\end{figure}

The latency in information exchange was compared to two alternatives, namely a laptop (HP Compaq Pro 6300 MT) and a Raspberry Pi (4th Gen Model B). Specifically, the round-trip delay of the communication was used as the key performance indicator. That is the combined duration of message transmission and acknowledgement with the network configuration depicted in \autoref{fig:com_delay_test}. For example, different devices (i.e., a smartphone, a laptop and a Raspberry Pi) sent a message to the laptop within the network of Empa (152.88.x.x), which acknowledged the receipt of the message. Each device quantified the time interval to receive the acknowledgement. In all cases, only the smartphone was connected to Wi-Fi, reducing communication performance.

    \section{Results}\label{sec:Results}
	The experiments utilized the NEST building located at the Empa campus in D\"ubendorf \cite{richner2018nest}, discarded smartphones,  a private network and the network of Empa. Two smartphones were used in the experiments with key specifications summarized in \autoref{tab:CaseStudies}.
\begin{center}
    \begin{table}[!htbp]
    \centering
    \caption{\label{tab:CaseStudies}List of mobile devices used in the experiments.} 
        \begin{tabular}{llllll}
        \toprule
        ID & CPU & RAM & WLAN & Platform & Release year\\
        \midrule
        1 & Octa-core (Kirin 970) & 6GB & Wi-Fi 802.11 & Android & 2018 \\
        2 & Octa-core (Exynos 7904) & 4GB & Wi-Fi 802.11 & Android & 2019 \\
        \bottomrule
        \end{tabular}
    \end{table}
\end{center}

\subsection{Applicability to control tasks}
\begin{figure}[htbp]
    \centering
    \includegraphics[width =.9\textwidth]{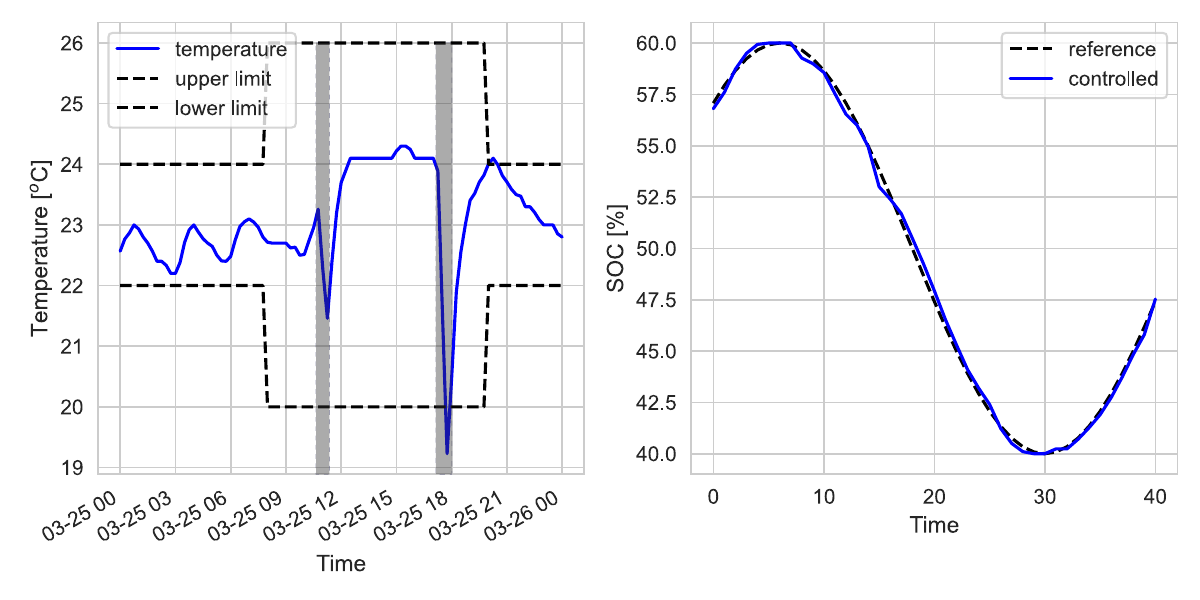}
    \caption{Experiment results of control-related case studies. The figure on the left shows room temperature control results, where the black dashed lines indicate the comfort limits and the blue line shows the realized temperature trajectory. The figure on the right shows battery reference tracking control results, where the black dashed line shows the reference signal and the blue line shows the measured \ac{SOC}.}\label{fig:control_cases}
\end{figure}

Python 3.8 was used together with Mosek \cite{mosek} as the solver. The results of control-related case studies are summarized in \autoref{fig:control_cases}. In \textit{Case 1}, it can be observed that the temperature was controlled within the predefined comfort zone most of the time. The shaded areas indicate periods of window opening, which led to substantial temperature drops. These occupants-induced constraint violations were beyond the capability of the heating system and the controller. Therefore, these instances do not suggest inapplicability.
In \textit{Case 2}, the mean tracking error and the root-mean-square-error were $-0.02$~\% and $0.09$~\%, respectively. The tracking errors can be attributed to the controller and the actuation precision. Both experiments show satisfactory results and confirm the applicability to control tasks.
\subsection{Applicability to communication tasks}
\begin{figure}[htbp]
    \centering
    \includegraphics[width =.7\textwidth]{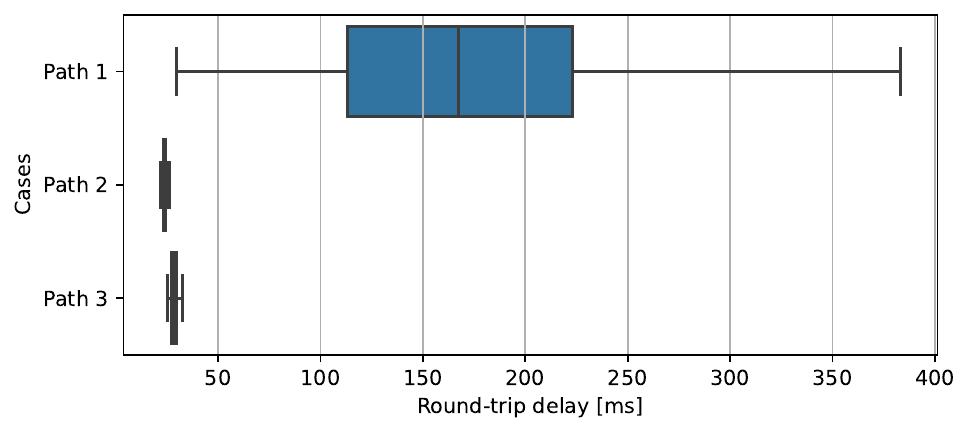}
    \caption{Boxplots of round-trip delays with various hardware and connection interfaces.}\label{fig:com_delay_results}
\end{figure}

The communication delays of all three paths depicted in \autoref{fig:com_delay_test} are quantified and compared in \autoref{fig:com_delay_results}. It can be observed that \textit{Path 1} exhibits the highest delays due to its internet connection using Wi-Fi. Nonetheless, the delays mostly fall within 1 second. Due to the large inertia of buildings, typical control time interval ranges from 30 minutes to hours. Therefore, communication delays within 1 second are considered acceptable for building-level energy management. However, such latency may be too large for tracking high-frequency signals when providing ancillary services to the power system. Potential solution includes adding peripherals to enable Ethernet connection for reduced latency and improved stability at the expense of increased cost and carbon footprint.

	\section{Conclusion}\label{sec:Conclusion}
	This paper experimentally evaluates the applicability of reusing discarded smartphones to typical control and communication tasks in buildings. Various controlled systems, control tasks, and algorithms have been considered. The results verify technical feasibility and reveal descent performance. As the capabilities of smartphones continue to improve over time, better-performing reusable resources can be expected. The proposed reuse adds to the existing portfolio of circularity concepts and sheds light on enhancing sustainability in the built environment.

Several limitations must be noted. First, the security of the chain of open-source tools has not been verified. This is crucial in future studies as the security of the system, which the distributed devices are integrated into, can be significantly affected. Second, the long-term stability, which industrial PLCs excel at, needs to be further examined. 
The next step includes achieving scalability through software standardization and investigating a simpler chain of tools to ensure stability, security, and efficiency. Additional life-cycle analysis can be performed in the future to quantify the impacts on sustainability. 

	\begin{acronym}
	\acro{CHP}{combined heat and power plant}
	\acro{COP}{coefficient of performance}
	\acro{DER}{distributed energy resource}
	\acro{DH}{district heating}
	\acro{DHN}{district heating network}
	\acro{DHO}{district heating operator}
	\acro{DR}{demand response}
	\acro{DSM}{demand side management}
	\acro{DSO}{distribution system operator}
	\acro{ENTSO-E}{European network of transmission system operators for electricity}
	\acro{EV}{electric vehicle}
	\acro{HP}{Heat Pump}
	\acro{ICT}{information and communications technology}
	\acro{MQTT}{message queuing telemetry transport}
	\acro{MPC}{model predictive control}
    \acro{OCP}{optimal control problem}
	\acro{PV}{photovoltaic}
	\acro{RES}{renewable energy sources}
    \acro{SOC}{state-of-charge}
	\acro{SES}{smart energy system}
	\acro{SH}{Space Heating}
    \acro{SMM-PC}{signal matrix model predictive control}
	\acro{TSO}{transmission system operator}
	\acro{V2G}{vehicle-to-grid}
    \acro{SSH}{secure shell protocol}
\end{acronym}

\section*{Acknowledgments}
We would like to thank Philipp Heer and Julie Rousseau for their insightful discussions and contribution of electronics used in this work. 
	\bibliographystyle{ieeetr}
    \bibliography{tail/library,tail/websites}
	
\end{document}